\documentclass[final,5p,times,twocolumn]{elsarticle}

\usepackage{physics}
\usepackage[english]{babel}
\usepackage[utf8]{inputenc}
\usepackage{graphicx,epstopdf,color}
\usepackage{xspace}
\usepackage{float}
\usepackage{booktabs}
\usepackage{hyperref}
\journal{Computational Materials Science}

\begin{document}

\begin{frontmatter}

\title{Parameterizing DFT+U+V from Hybrid Functionals: A Wannier-Function-Based Approach for Correlated-Subspace Hamiltonians}

\author[imp]{Dmitry M. Korotin\corref{cor1}}
\ead{dmitry@korotin.name}
\cortext[cor1]{Corresponding author}

\author[imp]{Anna A. Anisimova}

\author[imp,urfu]{Vladimir I. Anisimov}

\affiliation[imp]{
  organization={M.N. Mikheev Institute of Metal Physics, Ural Branch of the Russian Academy of Sciences},
  addressline={18 S. Kovalevskaya St.},
  city={Yekaterinburg},
  postcode={620108},
  country={Russia}
}

\affiliation[urfu]{
  organization={Department of Theoretical Physics and Applied Mathematics, Ural Federal University},
  addressline={19 Mira St.},
  city={Yekaterinburg},
  postcode={620002},
  country={Russia}
}

\begin{abstract}
We present an approach to parameterize DFT+$U$+$V$ from hybrid-functional 
calculations using Wannier-function projections. Semilocal and hybrid electronic
structures are represented using the same trial orbitals, energy windows, and orbital
ordering, and effective on-site ($U$) and intersite ($V$) parameters are obtained by
minimizing the Hamiltonian mismatch within the selected correlated subspace. We apply
the procedure to MgO, NiO, and V$_2$O$_5$, which represent wide-gap,
charge-transfer, and formally $d^0$ oxides, respectively. For all three systems, the
mapped correction brings the band gaps, densities of states, and magnetic moment closer
to the hybrid-functional reference than semilocal DFT. The fitted parameters are
representation-dependent effective quantities, and the workflow provides a reproducible
route for constructing DFT+$U$+$V$ models for a specified Wannier protocol.
\end{abstract}


\end{frontmatter}

\section{Introduction}

Local and semilocal exchange-correlation functionals often suffer from self-interaction error and inaccurate orbital-dependent level alignment, especially in systems containing localized atomic-like states. These deficiencies can lead to underestimated band gaps, incorrect relative positions of ligand and metal states, and inaccurate magnetic moments. 
The DFT+$U$ method partially remedies self-interaction and level-alignment errors by introducing a Hubbard-like correction for localized electrons~\cite{Anisimov1991a,Liechtenstein1995,Dudarev1998,Anisimov1997}. This correction penalizes fractional occupations of localized orbitals, thereby reducing the self-interaction error inherent in conventional DFT and opening band gaps in systems incorrectly predicted to be metallic. The inter-site $+V$ extension in DFT+$U$+$V$ further improves the description by incorporating inter-site Coulomb interactions, which are particularly important in systems with significant charge transfer, orbital hybridization, or covalent bonding~\cite{DFTUV}. Together, the $+U$ and $+V$ corrections enhance the accuracy of DFT~\cite{TimrovLiIon,TimrovSrTiO3} in describing charge localization, orbital ordering, magnetic properties, and band structures, while preserving its favorable computational scaling compared to more demanding many-body approaches such as GW or dynamical mean-field theory (DMFT)~\cite{Georges1996,Kotliar2006}.

A central challenge in applying DFT+$U$+$V$ is choosing the Hubbard parameters $U$
and $V$. These parameters are not fixed by the formalism and must be supplied externally.
Two common strategies are (i) empirical tuning to reproduce selected experimental
observables and (ii) first-principles estimates, for example from linear-response (LR)
theory~\cite{Cococcioni2005}, constrained DFT~\cite{cDFT}, constrained random phase
approximation (cRPA)~\cite{Aryasetiawan2004}.
Empirical tuning is straightforward when reliable data are available, but it has limited predictive power and becomes impractical for systems or conditions without experimental reference data. First-principles
approaches, such as LR and cRPA, are more systematic and transferable, but they can be computationally demanding
and sensitive to technical choices such as the basis set or supercell size. Moreover,
inter-site interactions ($V$) are often neglected because they are more difficult to
compute accurately.

Another line of work treats Hubbard parameters as effective quantities chosen to
reproduce either experiment or a higher-level electronic-structure reference. Early
DFT+$U$ studies of transition-metal oxides and oxide defects showed that a suitable
$U$ can recover localization and gap opening similar in spirit to hybrid-functional
corrections, although this correspondence is generally qualitative and depends on the
property considered~\cite{Morgan2007,ArroyoDeDompablo2011}. More recent approaches
make this calibration systematic by optimizing $U$ values against hybrid-functional
band structures using Bayesian optimization~\cite{Yu2020}, extending the same idea
to high-dimensional $U$ and $V$ spaces with active learning~\cite{Yu2023}, and using
machine learning to predict structure-dependent Hubbard parameters from such optimized
data~\cite{Cai2024}. These works show that fitted Hubbard parameters are effective,
material- and environment-dependent quantities rather than universal constants.

Hybrid exchange--correlation functionals provide an alternative way to improve the description of localized electrons by mixing a fraction of exact exchange with a semilocal functional and, in screened/range-separated variants, attenuating the long-range part of the Fock term~\cite{Heyd2003a,Heyd2006,Adamo1999,Becke1993}. In many transition-metal oxides, this reduces self-interaction and improves the relative energetic alignment of ligand $p$ and metal $d$ states, leading to more realistic band gaps and magnetic moments compared to semilocal DFT.

The main limitation of hybrid functionals is the computational cost associated with evaluating the nonlocal exchange operator, which can become prohibitive for large cells, dense $k$-point meshes, or geometry optimization. Consequently, hybrids are often used as a reference for the electronic structure, while more affordable corrected semilocal approaches (e.g., DFT+$U$+$V$) are preferred for extensive structural or configurational sampling.

The connection between hybrid functionals and DFT+$U$ corrections can be made more
explicit for localized orbitals. Ivady et al.~\cite{Ivady2014} showed that, when the hybrid-functional
exchange contribution is projected onto a localized atomic-like subspace, its
occupation-dependent part has the same form as the rotationally invariant Dudarev
DFT+$U$ correction~\cite{Dudarev1998}. In this correspondence, the effective
interaction strength is controlled by the hybrid mixing parameter and the on-site
Coulomb and exchange integrals. This provides a formal basis for interpreting DFT+$U$
as a localized, screened Hartree--Fock-like correction. In the present work we use the
same physical connection in the opposite direction.
The key idea of our approach is to construct an effective DFT+$U$+$V$
representation of a selected hybrid-functional electronic structure within a specified
localized subspace. The method proceeds in four steps:
\begin{enumerate}
    \item Calculate the electronic structure of the target compound using a hybrid functional.
    \item Transform the hybrid-functional Bloch states into a localized Wannier representation and extract the corresponding Hamiltonian matrix elements within the correlated subspace.
    \item Repeat the same procedure using a standard DFT functional (LDA/GGA), including the construction of the corresponding Wannier-function Hamiltonian.
    \item Map the hybrid-functional Hamiltonian onto the DFT Hamiltonian augmented with DFT+$U$+$V$ corrections, treating the interaction parameters as fitting variables.
\end{enumerate}

The use of a Wannier representation is important because it makes the mapping
basis-explicit. Similar Wannier-downfolding ideas have been used to compare PBE,
PBE+$U$, hybrid-functional, and partially self-consistent GW electronic structures by
extracting tight-binding parameters in a common low-energy basis~\cite{Franchini2012}.
Our construction follows this Hamiltonian-level idea, but with a different goal: instead
of fitting an unrestricted tight-binding model, we constrain the difference between the
semilocal and hybrid Wannier Hamiltonians to have the form generated by on-site $U$ and
inter-site $V$ corrections.

In this way, one obtains both (i) a set of localized basis functions [Wannier functions
(WFs)] suitable for describing the low-energy electronic states near the Fermi level and
(ii) an effective extended-Hubbard parametrization of the hybrid-functional correction,
including inter-site interactions for an arbitrary number of coordination shells. Because
the basis, energy window, and interaction range are explicit, the extracted parameters
can be interpreted not only as inputs for less expensive DFT+$U$+$V$ calculations for
related larger models, such as slabs, heterostructures, or defect supercells, but also
as effective on-site and inter-site interactions for more advanced electronic-structure
approaches such as DFT+DMFT~\cite{Georges1996,Kotliar2006} or for low-energy model
analyses, provided that the local chemical environment of the corrected orbitals remains
sufficiently similar. At the same
time, the mapping depends on these technical choices:
the Wannier projection, energy window, and real-space cutoff define the subspace in
which the parameters are fitted. The resulting $U$ and $V$ values should therefore be
regarded as effective quantities for this specified subspace, rather than unique
material constants.

It should be noted that self-consistent pseudohybrid Hubbard functionals provide an
already existing route to reduce the arbitrariness of Hubbard parameters. In the ACBN0
approach, the on-site Hubbard correction is evaluated from Hartree--Fock-like Coulomb
integrals and renormalized occupation matrices, so that the effective $U$ parameters are
determined self-consistently from the electronic structure rather than fitted
empirically~\cite{Agapito2015}; this strategy has been applied, for example, to improve
structural and electronic properties of perovskite oxides~\cite{MayKolpak2020}. The
same idea has also been generalized to inter-site interactions, leading to pseudohybrid
DFT+$U$+$V$ schemes in which both $U$ and $V$ are obtained self-consistently and can
substantially improve band gaps and lattice-dynamical properties of covalent
semiconductors~\cite{Lee2020a,Yang2021,Tancogne-Dejean2020}. These methods are
conceptually related to the present work because they also recast part of
screened-exchange physics into an extended-Hubbard form. However, their goal and
construction are different: ACBN0-type approaches define the Hubbard parameters
internally within a selected localized subspace and do not aim to reproduce a chosen
hybrid functional. In contrast, our approach uses the hybrid-functional Hamiltonian as
an explicit reference and fits effective $U$ and $V$ parameters so that the DFT+$U$+$V$
Hamiltonian reproduces it in a fixed Wannier basis. Thus, the similarity is mainly in
the use of an extended-Hubbard language, while the present method is a
hybrid-to-DFT+$U$+$V$ mapping procedure rather than a self-consistent pseudohybrid
functional.

In this work, we test the proposed workflow on three oxides with different
electronic structures: MgO, a wide-gap insulator without transition-metal ions
for which semilocal DFT still underestimates the band gap; NiO, an antiferromagnetic
charge-transfer insulator and a standard benchmark for correlation effects; and
V$_2$O$_5$, a transition-metal oxide with a formally empty $d$ shell ($d^0$) for which
the DFT band gap is also substantially underestimated. Thus, the benchmark set is not
restricted to strongly correlated systems: MgO and V$_2$O$_5$ test the ability of the
mapping to reproduce hybrid-functional level alignment and hybridization effects, while
NiO tests the method for a correlated charge-transfer insulator.

\section{Methods}

\subsection{Extraction of effective $U$ and $V$ parameters}

We construct WFs by projecting the pseudoatomic orbitals onto a selected subspace of
Bloch eigenstates and performing a L\"owdin orthonormalization independently at each
$\vb{k}$ point. A detailed description of this procedure within the
pseudopotential framework can be found in Ref.~\cite{hamilt}.

In reciprocal space, the projected WFs are
\begin{equation}
\label{Wannier:proj}
\ket{W_{n\vb{k}}} \equiv \sum_{\mu=N_1}^{N_2} \ket{\Psi_{\mu\vb{k}}} \braket{\Psi_{\mu\vb{k}}}{\phi_{n\vb{k}}}.
\end{equation}
Here, $\{\ket{\Psi_{\mu\vb{k}}}\}$ are Bloch eigenstates in the chosen band window
from $N_1$ to $N_2$. The resulting WFs are localized and retain the orbital
character of the trial functions.
The functions $\ket{\phi_{n\vb{k}}}$ are the Bloch sums of the corresponding
pseudoatomic orbitals.

Real-space WFs are obtained by the Fourier transform
\begin{equation}
\ket{W_{n}^{\vb{T}}} = \frac{1}{\sqrt{N_{\vb{k}}}} \sum_{\vb{k}} e^{-i\vb{k}\cdot\vb{T}} \ket{W_{n\vb{k}}},
\end{equation}
where $N_{\vb{k}}$ is the number of $\vb{k}$ points in the Brillouin-zone sampling. 

To avoid ambiguity in real space, we use the following notation: $i,j$ label atoms
in the reference unit cell, $\vb{T}$ is a Bravais-lattice translation vector, and
$I\equiv(i,\vb{T}_I)$ labels a site in the replicated supercell (cluster). Orbital
indices within the correlated shell on a given atom are denoted by $m,n$.

The matrix elements of the one-electron Hamiltonian in reciprocal space are defined as
\begin{equation}
\label{Wannier:Ham-k}
H_{mn}^{\sigma}(\vb{k}) =
\bra{W_{m\vb{k}}}
\left( \sum_{\mu=N_1}^{N_2} \ket{\Psi_{\mu\vb{k}}} 
\varepsilon_\mu^{\sigma}(\vb{k}) 
\bra{\Psi_{\mu\vb{k}}} \right)
\ket{W_{n\vb{k}}},
\end{equation}
where $\varepsilon_\mu^{\sigma}(\vb{k})$ is the eigenvalue of the one-electron Hamiltonian corresponding to band $\mu$ and spin $\sigma$.

The matrices $H^{\sigma}(\vb{k})$ are obtained in a post-processing step following
the self-consistent cycle using the same WF projection procedure for both conventional
DFT (denoted below as $H^{\mathrm{DFT}}$) and hybrid-functional calculations
(denoted as $H^{\mathrm{hyb}}$).

The spin-resolved occupation matrix in the WFs basis is defined as
\begin{equation}
\label{Wannier:Occ-k}
n_{mn}^{\sigma}(\vb{k}) =
\bra{W_{m\vb{k}}}
\left( \sum_{\mu=N_1}^{N_2} \ket{\Psi_{\mu\vb{k}}} 
f_\mu^{\sigma}(\vb{k}) 
\bra{\Psi_{\mu\vb{k}}} \right)
\ket{W_{n\vb{k}}},
\end{equation}
where $f_\mu^{\sigma}(\vb{k})$ is the occupation number (typically the Fermi--Dirac
weight) of the Bloch state $\mu$ with spin $\sigma$ and wavevector $\vb{k}$.

To determine the inter-site interaction parameters $V$, we require the Hamiltonian
and occupation matrices between WFs centered on different atomic
sites in real space. These quantities are obtained by Fourier transforming the
$\vb{k}$-dependent WF matrices to real space and retaining the
matrix elements corresponding to atoms located within a sphere of radius
$R_{\mathrm{cl}}$ centered on a chosen reference correlated atom. This procedure
defines an effective finite cluster in real space, without performing explicit
supercell calculations. We denote the corresponding set of retained sites by
$\mathcal{C}(R_{\mathrm{cl}})$; in the cluster matrices below, the site indices are
restricted to $I,J\in\mathcal{C}(R_{\mathrm{cl}})$.

For $X\in\{\mathrm{DFT},\mathrm{hyb}\}$ we define
\begin{align}
H^{X,\mathrm{cl}}_{mI,nJ,\sigma}
&=
\frac{1}{\sqrt{N_{\vb{k}}}} \sum_{\vb{k}} 
e^{-i\vb{k}\cdot(\vb{T}_J-\vb{T}_I)}\,
H^{X,\sigma}_{mn}(\vb{k}),
\nonumber\\
n^{X,\mathrm{cl}}_{mI,nJ,\sigma}
&=
\frac{1}{\sqrt{N_{\vb{k}}}} \sum_{\vb{k}} 
e^{-i\vb{k}\cdot(\vb{T}_J-\vb{T}_I)}\,
n^{X,\sigma}_{mn}(\vb{k}).
\label{eq:cluster_matrices}
\end{align}

This procedure truncates the fitted inter-site part of the effective model beyond
$R_{\mathrm{cl}}$. It does not imply that the screened exchange in the hybrid functional
is strictly local or nearest-neighbor; rather, only the Hamiltonian blocks retained in
$\mathcal{C}(R_{\mathrm{cl}})$ are represented by explicit $V_{IJ}$ parameters. In
principle, the mapping can be extended to any cluster size by increasing
$R_{\mathrm{cl}}$. In practice, however, the present Quantum ESPRESSO implementation of
the $+V$ correction can handle only a limited number of distinct inter-site
interactions in a tractable input setup. Therefore, retaining the first and, when
needed, second nearest-neighbor shells is a reasonable practical choice for subsequent
DFT+$U$+$V$ calculations, especially because these shells usually contain the strongest
hybridization matrix elements.

In particular, Eq.~\eqref{eq:cluster_matrices} gives $H^{\mathrm{DFT,cl}}$ and
$H^{\mathrm{hyb,cl}}$; the same notation is used for the occupation matrices
$n^{\mathrm{DFT,cl}}$ and $n^{\mathrm{hyb,cl}}$.

Following the rotationally invariant extended DFT+$U$+$V$ formalism~\cite{DFTUV},
within this cluster basis the DFT+$U$+$V$ correction to the one-electron
Hamiltonian is written as
\begin{align}
\Delta H^{\sigma}_{mI,nJ}
&=
\delta_{IJ}\, U_I
\left(
\frac{1}{2}\delta_{mn}
-
n^{\mathrm{DFT,cl}}_{mI,nI,\sigma}
\right)
\nonumber\\
&\quad -
(1-\delta_{IJ})\, V_{IJ}\,
n^{\mathrm{DFT,cl}}_{mI,nJ,\sigma}.
\label{eq:deltaH_UV}
\end{align}
Here, $U_I$ is the on-site Hubbard parameter on site $I$ and $V_{IJ}$ is the inter-site
interaction parameter for the retained pair of sites $(I,J)$. Increasing $R_{\mathrm{cl}}$ adds further
inter-site Hamiltonian blocks and possible $V_{IJ}$ parameters, whereas decreasing it
removes longer-range terms from the effective DFT+$U$+$V$ model Hamiltonian.

The resulting effective Hamiltonian is
\begin{equation}
H^{\mathrm{model}}_{\sigma}
=
H^{\mathrm{DFT,cl}}_{\sigma}
+
\Delta H^{\sigma}(U,V).
\end{equation}

The parameters $\{U_I\}$ and $\{V_{IJ}\}$ are determined by fitting
$H^{\mathrm{model}}_{\sigma}$ to the corresponding Hamiltonian
obtained from hybrid-functional calculations, $H^{\mathrm{hyb,cl}}_{\sigma}$.
We minimize the following cost function:
\begin{equation}
\mathcal{L}(U,V)
=
\sum_{\sigma}
\left\|
H^{\mathrm{hyb,cl}}_{\sigma}
-
\left(
H^{\mathrm{DFT,cl}}_{\sigma}
+
\Delta H^{\sigma}(U,V)
\right)
\right\|^2 ,
\label{eq:loss_function}
\end{equation}
where the norm corresponds to the sum of squared differences between all matrix elements.
Since $\Delta H^{\sigma}(U,V)$ depends linearly on the parameters, Eq.~\eqref{eq:loss_function} reduces to a standard linear least-squares problem.
In the present implementation, this least-squares metric is evaluated over the
independent upper-triangular matrix elements of the Hermitian Hamiltonian
matrices; the omitted lower-triangular elements are redundant by Hermiticity.
This convention defines a metric over the independent matrix elements rather
than the full Frobenius norm, for which off-diagonal elements would carry twice
the weight. With this convention, Eq.~\eqref{eq:loss_function} still fits the
Wannier Hamiltonian matrix elements directly, rather than only the eigenvalue
spectrum. It therefore constrains both diagonal terms (level alignment and gaps)
and off-diagonal terms (hybridization and effective hoppings) within the
correlated subspace.
As a result, the mapped parameters are better suited for subsequent
structural relaxations and many-body extensions, where the full Hamiltonian matrix is
required.
For the systems considered here, we verified that the semilocal and hybrid-functional occupation matrices in the fitted subspace differ only moderately; the corresponding occupation-matrix differences and integrated PDOS weights are reported in the Supplemental Material.

In the calculations presented here,
each on-site parameter is assigned to an atom and angular-momentum shell, for example
Mg-$s$, Mg-$p$, O-$p$, or Ni-$d$. The inter-site parameters are assigned to pairs of such
shells on different sites within the cluster cutoff. Thus, the number of fitted variables
is the number of retained on-site shell blocks plus the number of retained inter-site
block pairs. Symmetry-equivalent atoms and bonds are treated in the same way when they
are equivalent in the chosen structure and Wannier basis, while inequivalent bonds are
listed separately. In principle, the same mapping could be generalized to fit full
orbital-resolved $U$ and $V$ matrices. In the present work we use shell-resolved
parameters because this is the form that can be used directly in the current Quantum
ESPRESSO implementation of DFT+$U$+$V$.

The fitted parameters are defined for a specific correlated subspace. In this work, we
construct this subspace using projection WFs based on the same pseudoatomic trial
orbitals for the semilocal and hybrid calculations. This procedure keeps the WFs close
to the atomic-like projectors used in the subsequent DFT+$U$+$V$ step and helps to
preserve the correspondence between orbitals in the two calculations. The energy window,
gauge, and orbital ordering must therefore be chosen consistently. A different unitary
rotation within the selected subspace would generally change individual matrix elements.
Permutations of equivalent orbitals do not affect the mapping if the same ordering is
used for both Hamiltonians.

The projection procedure, however, does not guarantee that the semilocal and hybrid WFs
are identical because the corresponding Bloch subspaces may differ. This basis mismatch
is a non-ideality of the method and introduces uncertainty in the extracted parameters.
The mapping should therefore be used only when the same projection procedure produces
sufficiently similar target subspaces for both functionals.

For systems with entangled bands, the method relies on the ability of the selected
projection window to define a stable target orbital subspace. If bands outside the
intended manifold strongly mix with the target states, different windows or
disentanglement choices may lead to different effective parameters. In such cases the
mapping remains well defined for a specified projection protocol, but the robustness of
the resulting $U$ and $V$ values must be checked explicitly.

\subsection{Details of ground-state calculations}

The electronic structure calculations were performed using the \textsc{Quantum ESPRESSO}
package~\cite{Giannozzi2009,giannozzi2020}. Pseudopotentials were taken from the Standard
Solid-State Pseudopotentials (SSSP) library, using the PBEsol Efficiency set
(v1.3.0)~\cite{prandini2018} and the kinetic-energy cutoffs recommended by the authors
for each element. Brillouin-zone integration was carried out on an $8 \times 8 \times 8$
Monkhorst--Pack $k$-point mesh in the irreducible Brillouin zone.
Hybrid-functional calculations used the HSE06 screened-exchange form~\cite{Heyd2003a}
with exact-exchange fraction $\alpha=0.25$, applied on top of the same PBEsol-based
setup, including the same pseudopotentials, structural model, and $\vb{k}$-point mesh.
Thus, the fitted parameters map the Hamiltonian difference between the semilocal PBEsol
calculation and the corresponding screened-hybrid reference.
The auxiliary $\vb{q}$-point grid used to evaluate the Fock exchange was set to $2 \times 2 \times 2$ for MgO, $4 \times 4 \times 4$ for NiO, and $2 \times 1 \times 4$ for V$_2$O$_5$.
The reported hybrid-functional gaps and the corresponding mapped $U$ and $V$ values should therefore be interpreted as results for this specified computational setup. Improving the convergence of the underlying hybrid-functional reference, for example by using denser Fock $\vb{q}$-point grids or different pseudopotentials, may change the numerical values of the extracted parameters but does not change the mapping procedure itself.
The crystal structures used in the calculations are described in the Supplemental Material.

\section{Results}
\subsection{MgO}

\begin{figure}[h]
    \includegraphics[width=\linewidth]{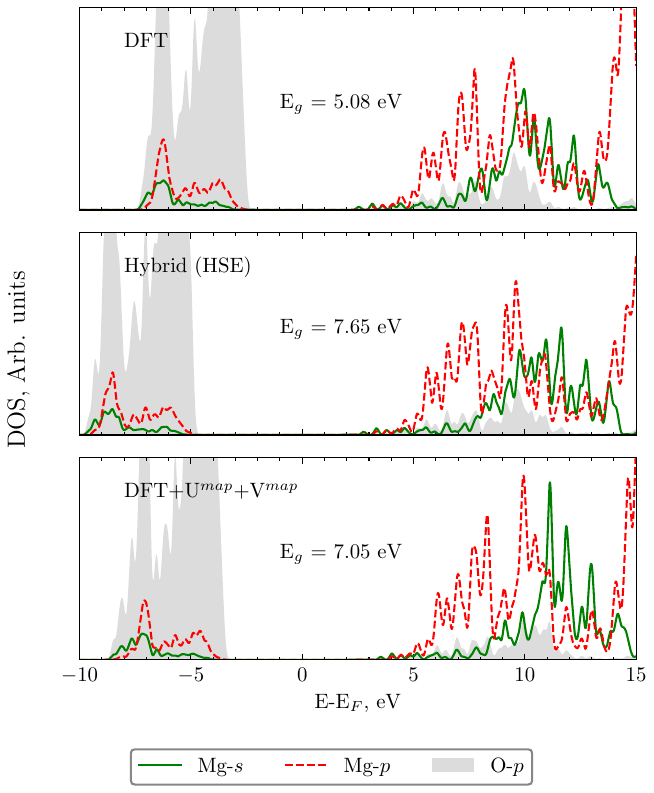}
    \caption{Projected density of states of MgO calculated using DFT (PBEsol), the HSE06 hybrid functional, and DFT+$U$+$V$ with parameters fitted from HSE06.}
    \label{fig:MgO_pdoses}
\end{figure}

Magnesium oxide crystallizes in the rock-salt structure~\cite{MgO_Structure} and is a
wide-gap insulator with an experimental gap of about 7.83~eV~\cite{whited1973exciton}.
Even though it has no partially filled $d$ or $f$ shells, MgO is a useful benchmark
because semilocal DFT still underestimates its gap~\cite{Yang2023,Gromann2026}.
Using PBEsol, we obtained a band gap of 5.08~eV, while HSE06 yields 7.65~eV. We did not
aim to reproduce the exact experimental gap; instead, we used the standard HSE06
parameters without tuning to test the proposed method.
The DFT partial densities of states (PDOS) in the top panel of Fig.~\ref{fig:MgO_pdoses}
show that states around the Fermi level have Mg $s$/$p$ and O $p$ character, consistent
with appreciable Mg--O hybridization.
Accordingly, we constructed Wannier functions for the Mg $s$ and $p$ states and for the O $p$ states, spanning an energy window from $-8$~eV to $20$~eV relative to the Fermi level.

The mapping yielded the following on-site parameters: $U_{\mathrm{Mg}\,s} = 1.05$~eV,
$U_{\mathrm{Mg}\,p} = 0.44$~eV, and $U_{\mathrm{O}\,p} = 5.80$~eV. For the inter-site
terms, we obtained $V_{\mathrm{Mg}\,s - \mathrm{O}\,p} = 1.96$~eV and
$V_{\mathrm{Mg}\,p - \mathrm{O}\,p} = 1.98$~eV.

Figure~\ref{fig:MgO_pdoses} compares the PDOS obtained using DFT (PBEsol), the HSE06 hybrid
functional, and DFT+$U$+$V$ with the mapped parameters. The DFT+$U$+$V$ calculation
increases the band gap to 7.05~eV, bringing it closer to both the HSE06 result and
experiment. Moreover, the overall PDOS shape obtained within DFT+$U$+$V$ closely follows
that of the hybrid-functional calculation. A quantitative comparison based on
orbital-resolved PDOS centers of gravity below and above the Fermi level is
provided in the Supplemental Material.

An exact match between the DFT+$U$+$V$ and HSE06 band gaps is not expected. One reason is
that the correlated subspace differs between the two approaches. In the hybrid-functional
calculation, all valence and conduction states are treated on an equal footing, whereas
in DFT+$U$+$V$ only a selected subset of states is explicitly corrected. In addition, for
simplicity, we included only the first shell of nearest neighbors when evaluating the
inter-site parameters $V$ for MgO. These differences lead to small residual
discrepancies in the resulting band gaps.

Large Hubbard interactions on oxygen $p$ orbitals are not merely a methodological artifact. 
Recent estimates based on the constrained random-phase approximation (cRPA) and
related schemes report effective $U_{pp}$ values of several electron-volts (typically
$5$--$8$~eV)~\cite{Ezeakunne2025,Outerovitch2023} for a broad class of oxides, including
both Mott--Hubbard and charge-transfer insulators. These values are comparable to typical
$U_{dd}$ or $U_{ff}$ parameters and reflect the finite localization of oxygen $p$ states
and incomplete screening in oxides, rather than an ``overcorrection'' for nominally
filled bands.

As a simple structural check, we also used the mapped parameters in equation-of-state (EOS) calculations based on isotropic volume scaling. For MgO, DFT gives $V_0=9.37$~\AA$^3$/ion and $B_0=161.8$~GPa, while DFT+$U$+$V$ gives $V_0=9.02$~\AA$^3$/ion and $B_0=178.1$~GPa; the experimental values are $V_0=9.34$~\AA$^3$/ion and $B_0=160.2$~GPa~\cite{Speziale2001MgO}. The full comparison is given in the Supplemental Material.

\subsection{NiO}

Nickel oxide is a prototypical strongly correlated material with antiferromagnetic (AFM)
order. It is a charge-transfer insulator where the band gap opens between oxygen $p$
states and nickel $d$ states~\cite{Sawatzky1984}. 
NiO is therefore a standard benchmark for beyond-DFT methods used to improve band gaps and magnetic moments compared with semilocal DFT~\cite{Anisimov1991a,Kunes2007,Bengone2000,DFTUV,DelloStritto2023}.

\begin{figure}[h]
    \includegraphics[width=\linewidth,height=0.70\textheight,keepaspectratio]{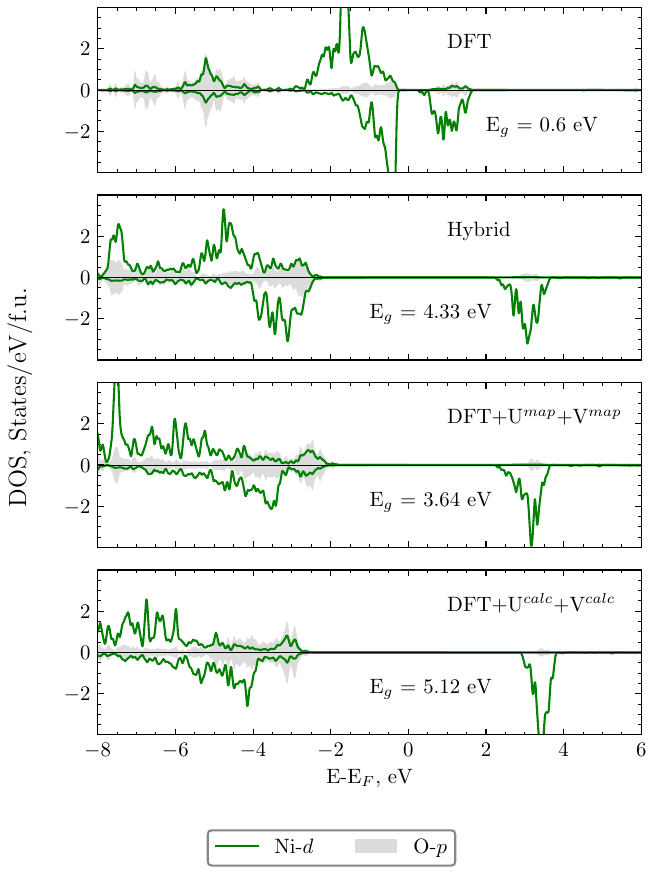}
    \caption{Projected density of states of NiO calculated using DFT (PBEsol), the HSE06 hybrid functional, DFT+$U^{map}$+$V^{map}$ with parameters mapped from HSE06, and DFT+$U^{calc}$+$V^{calc}$ with parameters from linear-response theory. For the antiferromagnetic state, the positive and negative branches show the spin-majority and spin-minority projections for a chosen Ni sublattice, respectively.}
    \label{fig:NiO_pdoses}
\end{figure}

The PDOS of NiO calculated using DFT (PBEsol) and HSE06 are shown in the top two panels of
Fig.~\ref{fig:NiO_pdoses}. PBEsol yields a band gap of 0.6~eV, which strongly
underestimates the experimental value of 3.7--4.3~eV~\cite{Powell1970,Sawatzky1984}. In
contrast, HSE06 predicts a band gap of 4.33~eV, in agreement with experiment.

To construct the correlated subspace for NiO, we included the Ni $3d$ and O $2p$ states,
which are essential for capturing the charge-transfer nature of the band gap. The
Wannier functions were generated by projection onto these atomic-like orbitals within an
energy window from $-8$~eV to $5$~eV relative to the Fermi level.

The mapping yielded $U_{\mathrm{Ni}\,3d} = 6.75$~eV and $U_{\mathrm{O}\,2p} = 2.64$~eV
for the on-site interactions, and $V_{\mathrm{Ni}\,3d - \mathrm{O}\,2p} = 1.54$~eV and
$V_{\mathrm{O}\,2p - \mathrm{O}\,2p} = 1.84$~eV for the inter-site interactions.
The third panel in Fig.~\ref{fig:NiO_pdoses} shows the PDOS obtained using DFT+$U$+$V$
with the mapped parameters. The calculated band gap is 3.64~eV, close to the HSE06
reference and in good agreement with experiment. The overall PDOS shape also aligns well
with that of the hybrid-functional calculation.

We also computed the extended-Hubbard parameters for NiO using linear-response
theory~\cite{Timrov2022a}. The obtained values are $U_{\mathrm{Ni}\,3d} = 7.61$~eV,
$U_{\mathrm{O}\,2p} = 9.69$~eV, and $V_{\mathrm{Ni}\,3d - \mathrm{O}\,2p} = 0.19$~eV.
These are consistent with the values reported in Ref.~\cite{Garcia-Uc2026}.
The PDOS obtained using DFT+$U$+$V$ with these parameters is shown in the bottom panel of
Fig.~\ref{fig:NiO_pdoses}. The calculated band gap is 5.12~eV, which overestimates both
the HSE06 result and experiment. This discrepancy is consistent with the larger $U$
values for the Ni $d$ and O $2p$ states obtained from linear response compared to those
derived from the mapping procedure.

In addition to the band gap, we compare the Ni magnetic moment estimated from L\"owdin
charges for the four approaches considered here (Table~\ref{tab:NiO_moment}). The PBEsol
value (1.32 $\mu_B$) is substantially smaller than that obtained with HSE06
(1.71 $\mu_B$). The DFT+$U^{map}$+$V^{map}$ result (1.69 $\mu_B$) is close to the
hybrid-functional reference, whereas DFT+$U^{calc}$+$V^{calc}$ yields a slightly larger
moment (1.83 $\mu_B$), consistent with the overestimation of the band gap in that setup.

\begin{table}[h]
    \centering
    \caption{Ni magnetic moment obtained from L\"owdin charges for NiO calculated using PBEsol, HSE06, DFT+$U^{map}$+$V^{map}$ with parameters mapped from HSE06, and DFT+$U^{calc}$+$V^{calc}$ with parameters from linear-response theory.}
    \label{tab:NiO_moment}
    \begin{tabular}{l|c}
        \hline \hline
        Method & $\mu_{\textrm{Ni}}$ ($\mu_B$) \\
        \hline
        PBEsol & 1.32 \\
        HSE06 & 1.71 \\
        DFT+$U^{map}$+$V^{map}$ & 1.69 \\
        DFT+$U^{calc}$+$V^{calc}$ & 1.83 \\
        \hline \hline
    \end{tabular}
\end{table}

All approaches considered here slightly underestimate the experimental Ni magnetic moment
of $1.90\,\mu_{\mathrm{B}}$~\cite{Cheetham1983}. This residual discrepancy can be attributed to
(i) the finite Wannier projection window, (ii) the absence of dynamical correlations
beyond static DFT+$U$+$V$ or hybrid exchange, and (iii) the sensitivity of local moments
to the balance between $U$ and $V$, which controls the degree of $p$--$d$ hybridization.
Nevertheless, the mapped DFT+$U$+$V$ parameters reproduce the hybrid-functional moment
closely, indicating that spin polarization is captured within the chosen low-energy subspace.

The same EOS test gives a useful structural check for NiO. DFT gives $V_0=8.74$~\AA$^3$/ion and $B_0=223.8$~GPa, while DFT+$U$+$V$ gives $V_0=9.14$~\AA$^3$/ion and $B_0=204.2$~GPa; the experimental values are $V_0=9.12$~\AA$^3$/ion and $B_0=192$~GPa~\cite{Gavriliuk2023NiO}. Thus, in NiO the mapped correction improves both the electronic properties and the simple EOS estimate. 

\subsection{V$_2$O$_5$}

Vanadium pentoxide (V$_2$O$_5$) is a layered transition-metal oxide that is often
described as a $d^0$ system with nominally empty V $3d$ states. Because vanadium is in a
high oxidation state, the hybridization between V-$d$ and O-$p$ states is substantial.
Experimentally, the band gap of $\alpha$-V$_2$O$_5$ is 2.5~eV~\cite{Bodo1967}. Despite the
absence of partially filled $d$ shells, semilocal DFT typically underestimates the gap
and can misplace the V $3d$ conduction-band manifold relative to the O $2p$ valence
states. For this reason, additional treatments of Coulomb interactions are often used to
describe V$_2$O$_5$ properties~\cite{Roginskii2021,Plotnikov2026}.
This makes V$_2$O$_5$ a useful test case for the present mapping strategy: it probes
whether the fitted $U$ and $V$ parameters can reproduce a hybrid-functional reference
even when correlation effects are not primarily of the Mott--Hubbard type.

Following the same procedure as for MgO and NiO, we first computed the electronic
structure of V$_2$O$_5$ using PBEsol and HSE06. PBEsol underestimates the band gap
(1.83~eV), while HSE06 yields a larger (overestimated) value of 3.43~eV. The hybrid gap
could be tuned by changing the fraction of exact exchange, but we did not pursue this
here because our goal is to validate the mapping approach.

To construct the correlated subspace, we included the V $3d$ and O $2p$ orbitals, which
provide a compact description of the states near the valence-band maximum and the
conduction-band minimum. We generated Wannier functions by projection onto these
atomic-like orbitals within an energy window that captures the full O $2p$ valence
manifold and the low-lying V $3d$ conduction states, from -7~eV to 6~eV relative to the
Fermi level.

\begin{figure}[h]
    \includegraphics[width=\linewidth,height=0.70\textheight,keepaspectratio]{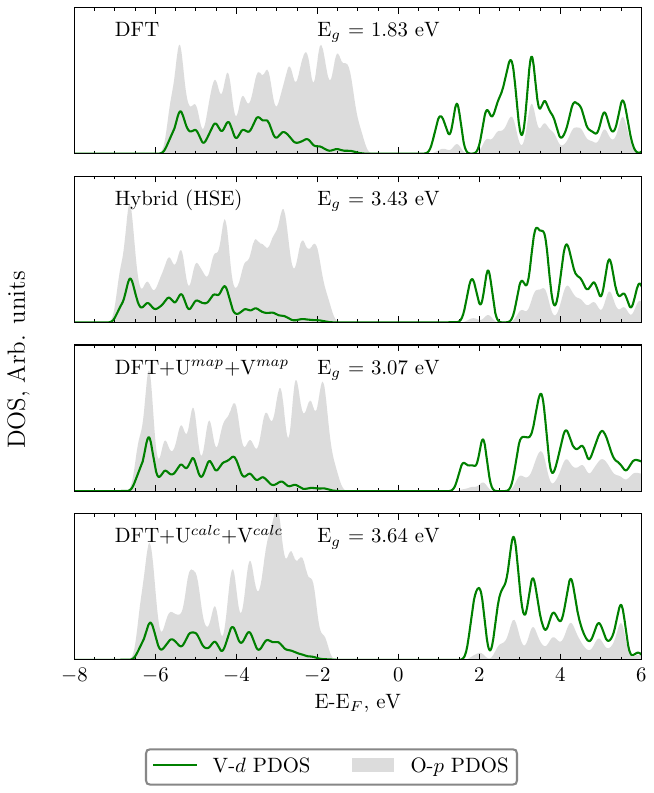}
    \caption{Projected density of states of V$_2$O$_5$ calculated using DFT (PBEsol), the HSE06 hybrid functional, DFT+$U^{map}$+$V^{map}$ with parameters mapped from HSE06, and DFT+$U^{calc}$+$V^{calc}$ with parameters from linear-response theory. The green and gray curves correspond to the V-$d$ and O-$p$ contributions, respectively.}
    \label{fig:V2O5_pdoses}
\end{figure}

The mapping yields the parameters summarized in Table~\ref{tab:V2O5_UV}, together with
the linear-response values used in the DFT+$U^{calc}$+$V^{calc}$ calculations. V$_2$O$_5$
contains three crystallographically inequivalent oxygen sites. Repeated oxygen labels in
the inter-site rows of the table denote distinct bonds involving the same oxygen type but
different neighboring atoms or bond lengths; the corresponding distances are listed
explicitly in the last column.

\begin{table*}[t]
    \centering
    \caption{Mapped (from HSE06) and linear-response (LR) interaction parameters for V$_2$O$_5$. The distance column is given only for inter-site terms; repeated oxygen labels in inter-site rows correspond to distinct bonds involving the same crystallographic oxygen type.}
    \label{tab:V2O5_UV}
    \begin{tabular}{l|c|c|c}
        \hline \hline
        Parameter & Mapped (eV) & LR (eV) & Inter-site distance (\AA) \\
        \hline
        $U_{\mathrm{V}\,3d}$ & 2.19 & 5.54 & -- \\
        $U_{\mathrm{O1}\,2p}$ & 2.62 & 7.57 & -- \\
        $U_{\mathrm{O2}\,2p}$ & 2.64 & 7.61 & -- \\
        $U_{\mathrm{O3}\,2p}$ & 2.40 & 7.22 & -- \\
        \hline
        $V_{\mathrm{V}\,3d - \mathrm{O1}\,2p}$ & 1.93 & 0.67 & 1.58 \\
        $V_{\mathrm{V}\,3d - \mathrm{O2}\,2p}$ & 1.80 & 1.01 & 1.88 \\
        $V_{\mathrm{V}\,3d - \mathrm{O2}\,2p}$ & 1.86 & 1.01 & 2.02 \\
        $V_{\mathrm{V}\,3d - \mathrm{O3}\,2p}$ & 1.83 & 0.95 & 1.78 \\
        $V_{\mathrm{O1}\,2p - \mathrm{O2}\,2p}$ & 1.10 & 0.59 & 2.74 \\
        $V_{\mathrm{O1}\,2p - \mathrm{O2}\,2p}$ & 1.13 & 0.59 & 2.87 \\
        $V_{\mathrm{O1}\,2p - \mathrm{O3}\,2p}$ & 1.11 & 0.53 & 2.65 \\
        $V_{\mathrm{O3}\,2p - \mathrm{O2}\,2p}$ & 1.29 & 0.68 & 2.74 \\
        $V_{\mathrm{O3}\,2p - \mathrm{O2}\,2p}$ & 0.81 & 0.68 & 3.67 \\
        $V_{\mathrm{O2}\,2p - \mathrm{O2}\,2p}$ & 1.61 & -- & 2.39 \\
        \hline \hline
    \end{tabular}
\end{table*}

Table~\ref{tab:V2O5_UV} shows that the two parameterization strategies yield
systematically different interaction strengths. The on-site terms obtained from linear
response are substantially larger than the mapped values (by roughly a factor of 2--3
for both V-$3d$ and O-$2p$). The inter-site terms show a less uniform trend: the mapped
V--O interactions are generally larger than their linear-response counterparts. For
O--O interactions, the two approaches yield values of the same order of magnitude (where
both are available), and these remain below the on-site values in both approaches. In
the mapped dataset, the inter-site interactions tend to decrease with increasing
interatomic separation, consistent with the finite range of the dominant hybridization
paths.

The variation of the mapped oxygen $U$ values should not be interpreted as a variation
of a universal atomic interaction. Calculations for oxides show that $U_{pp}$ can reach
several electron-volts and that its value and physical effect depend strongly on the
definition of the correlated orbitals and on the screening channels included in the
model~\cite{Outerovitch2023}. In the present mapping, $U_{\mathrm{O}\,2p}$ describes only
the on-site part of the PBEsol--HSE06 Hamiltonian difference within the selected Wannier
subspace. The larger value obtained for MgO suggests that this difference is mainly
represented by an on-site shift of the O-$2p$ states. In NiO and V$_2$O$_5$, a larger
part of the correction is distributed among the explicit $p$--$d$ and O--O inter-site
terms, resulting in smaller mapped oxygen $U$ values. Such a redistribution between
on-site and inter-site interactions is consistent with the extended DFT+$U$+$V$
framework, where $V$ describes hybridization and covalent bonding
effects~\cite{DFTUV,Lee2020a}.

Figure~\ref{fig:V2O5_pdoses} summarizes the PDOS obtained with semilocal DFT (PBEsol), the
HSE06 hybrid functional, and DFT+$U$+$V$ calculations. In agreement with the formal
$d^0$ configuration, the valence-band manifold is dominated by O-$p$ states, whereas the
conduction-band edge has strong V-$d$ character.

With the mapped parameters, the DFT+$U^{map}$+$V^{map}$ calculation opens the band gap to
3.07~eV and reproduces the main features of the hybrid-functional PDOS. For comparison,
DFT+$U^{calc}$+$V^{calc}$ yields a gap of 3.64~eV, which is slightly larger than the
HSE06 reference and deviates further from the experimental value of 2.5~eV.

We also applied the same EOS test to V$_2$O$_5$. DFT gives $V_0=12.94$~\AA$^3$/ion and $B_0=176.2$~GPa, while DFT+$U$+$V$ gives $V_0=13.01$~\AA$^3$/ion and $B_0=176.3$~GPa; the experimental values are $V_0=12.80$~\AA$^3$/ion and $B_0=50$~GPa~\cite{Loa2001V2O5HighPressure}. The large difference in $B_0$ reflects the limited isotropic protocol for this layered oxide, rather than a full structural optimization. 

\section{Residual of the Hamiltonian fit}

The quality of the mapping can be assessed directly at the Hamiltonian level from the
value of the least-squares objective in Eq.~\eqref{eq:loss_function}.  We denote by
$\mathcal{L}(0,0)$ the bare mismatch between the PBEsol and HSE06 Wannier Hamiltonians,
i.e., the value of the objective before adding the fitted DFT+$U$+$V$ correction, and by
$\mathcal{L}_{\min}$ the minimized value obtained with the fitted parameters.  Since
$\mathcal{L}$ itself depends on the number of orbitals, spin channels, and cluster
matrix elements included in the fit, we report the normalized root-mean-square residual
per independent Hamiltonian matrix element,
\begin{equation}
\Delta_{\mathrm{RMS}}(U,V)
=
\sqrt{\frac{\mathcal{L}(U,V)}{N_{\mathrm{elem}}}},
\end{equation}
where $N_{\mathrm{elem}}$ is the number of independent matrix elements included in the
least-squares fit, using the same upper-triangular convention as in
Eq.~\eqref{eq:loss_function}.  The fraction of the original squared mismatch removed by
the fitted correction is defined as
\begin{equation}
\eta
=
1 - \frac{\mathcal{L}_{\min}}{\mathcal{L}(0,0)} .
\end{equation}

\begin{table}[ht]
    \centering
    \caption{Normalized Hamiltonian-level residuals for the mapping from PBEsol to
    HSE06.  The quantity $\Delta_{\mathrm{RMS}}$ is the RMS residual per independent
    Hamiltonian matrix element, and the reduction $\eta$ is computed from the squared
    losses as $1-\mathcal{L}_{\min}/\mathcal{L}(0,0)$.}
    \label{tab:fit_residuals}
    \begin{tabular}{lccc}
        \hline \hline
        System &
        $\Delta_{\mathrm{RMS}}(0,0)$ (eV) &
        $\Delta_{\mathrm{RMS}}^{\mathrm{fit}}$ (eV) &
        $\eta$ (\%) \\
        \hline
        MgO & 0.711 & 0.003 & 99.9 \\
        NiO & 0.874 & 0.165 & 96.4 \\
        V$_2$O$_5$ & 0.276 & 0.028 & 98.9 \\
        \hline \hline
    \end{tabular}
\end{table}

Table~\ref{tab:fit_residuals} shows that the fitted DFT+$U$+$V$ correction removes more
than 96\% of the initial PBEsol--HSE06 Hamiltonian mismatch in all three test systems,
with the largest remaining RMS residual found for NiO.

\section{Conclusion}

We introduce a Wannier-function-based scheme to extract DFT+$U$+$V$ parameters from a
hybrid-functional reference by matching one-electron Hamiltonians in consistently
constructed localized representations. The on-site and inter-site interaction parameters are obtained by
minimizing the difference between the hybrid-functional Hamiltonian and a DFT Hamiltonian
supplemented with $U$ and $V$ corrections within a selected Wannier subspace. The
resulting parameters are effective quantities that depend on the chosen Wannier
representation rather than unique, universally transferable interaction strengths.

An important practical benefit is that the approach provides an explicitly defined
localized representation and a consistent set of extended Hubbard parameters for that
representation. Because the mapping is performed on the Hamiltonian matrix, it
constrains both diagonal level shifts and off-diagonal elements associated with orbital
hybridization and inter-site coupling.

We tested the method on three representative oxides—MgO, NiO, and V$_2$O$_5$. In all three
cases, the mapped parameters lead to DFT+$U$+$V$ electronic structures that are much
closer to the hybrid-functional results than standard semilocal DFT. In MgO, the
correction mainly shifts the O-$2p$ states relative to Mg-derived states, resulting in a
band gap close to that of HSE06. In NiO, treating both Ni-$3d$ and O-$2p$ states within
the correlated subspace reproduces the charge-transfer gap and the magnetic moment of
the hybrid-functional reference. In V$_2$O$_5$, where the states are strongly hybridized
despite the formal $d^0$ configuration, the mapping reduces the on-site $U$ values but
increases inter-site $V$ terms, leading to better agreement with the hybrid-functional
density of states than is obtained using the linear-response parameters.

The extracted $U$ and $V$ values depend on the Wannier energy window, the projection
scheme, and the real-space cutoff used to define the interaction cluster. This dependence
is inherent to any low-energy mapping because different subspaces capture different
aspects of the electronic structure. Thus, the reported parameters should be interpreted
as effective mapping parameters for the specified Wannier construction and interaction
cluster, rather than as unique material constants. Once this protocol is fixed, the
quality of the mapping can be assessed directly from the Hamiltonian-level residual. As an additional test, the same mapped parameters were used in inexpensive EOS calculations. For NiO, DFT+$U$+$V$ improves both the equilibrium volume and the bulk modulus relative to experiment, consistent with the recent all-electron DFT+$U$+$V$ study of Ref.~\cite{Beida2025DFTUVFLAPW}. For MgO, where semilocal DFT already describes the EOS well, no similar improvement is found.

The method provides a way to transfer hybrid-functional information obtained for a
representative cell into a semilocal DFT+$U$+$V$ framework. This can be useful for related
larger models, such as slabs, heterostructures, or defect supercells, where direct hybrid
calculations are too expensive, provided that the fitted parameters remain transferable.

\section*{CRediT authorship contribution statement}
\textbf{Dmitry M. Korotin:} Methodology, Software, Validation, Formal analysis,
Investigation, Data curation, Visualization, Writing -- original draft.
\textbf{Anna A. Anisimova:} Validation, Formal analysis, Investigation, Visualization.
\textbf{Vladimir I. Anisimov:} Conceptualization, Supervision, Project administration,
Funding acquisition, Resources.

\section*{Declaration of competing interest}
The authors declare that they have no known competing financial interests or personal
relationships that could have appeared to influence the work reported in this paper.

\section*{Declaration of generative AI and AI-assisted technologies in the writing process}
During the preparation of this work, the authors used OpenAI Codex for language editing
and preparation of submission materials. After using this tool, the authors reviewed
and edited the content as needed and take full responsibility for the content of the
publication.

\section*{Acknowledgments}
The Hubbard model parameter mapping procedure was developed with support from the
Russian Science Foundation (Project 24-12-00024). The linear-response calculations
were supported by the Ministry of Science and Higher Education of the Russian
Federation.

\section*{Data availability}
The Python code used to fit the Wannier-function Hamiltonian matrices and extract the
effective Hubbard parameters, together with example input and output files, is available
from Zenodo~\cite{Korotin2026Zenodo}. The current development version of the code is
also available at \url{https://github.com/dkorotin/UVextract}.
All calculations were performed using publicly available electronic-structure codes, and
the computational parameters required to reproduce the results are provided in the
Methods section and Supplemental Material.

\bibliographystyle{elsarticle-num}
\bibliography{main}

\end{document}